\def\mearth{{\rm\,M_\oplus}}

\documentclass[12pt]{article}

\usepackage{scicite}

\usepackage[pdftex]{graphicx} 

\usepackage{times}

\newenvironment{sciabstract}{%
\begin{quote} \bf}
{\end{quote}}

\renewcommand\refname{References and Notes}

\newcounter{lastnote}
\newenvironment{scilastnote}{%
\setcounter{lastnote}{\value{enumiv}}%
\addtocounter{lastnote}{+1}%
\begin{list}%
{\arabic{lastnote}.}
{\setlength{\leftmargin}{.22in}}
{\setlength{\labelsep}{.5em}}}
{\end{list}}

\title{The Empty Primordial Asteroid Belt}

\author
{Sean N. Raymond$^{1\ast}$, Andre Izidoro$^{1,2}$\\
\\
\normalsize{$^{1}$Laboratoire d'Astrophysique de Bordeaux,}\\
\normalsize{Univ. Bordeaux, CNRS, B18N, all{\'e}e Geoffroy Saint-Hilaire,}\\
\normalsize{33615 Pessac, France}\\
\normalsize{$^{2}$UNESP, Univ. Estadual Paulista,}\\
\normalsize{Grupo de Din{\`a}mica Orbital Planetologia,}\\
\normalsize{Guaratinguet{\`a}, CEP 12.516-410, S{\~a}o Paulo, Brazil}\\
\\
\normalsize{$^\ast$To whom correspondence should be addressed; E-mail:  sean.raymond@u-bordeaux.fr.}
}

\date{{\it Science Advances}, 3, e1701138 (2017)}

\begin{document} 


\baselineskip24pt


\maketitle


\begin{sciabstract}
The asteroid belt contains less than a thousandth of Earth's mass and is radially segregated, with S-types dominating the inner belt and C-types the outer belt. It is generally assumed that the belt formed with far more mass and was later strongly depleted. Here we show that the present-day asteroid belt is consistent with having formed empty, without any planetesimals between Mars and Jupiter's present-day orbits. This is consistent with models in which drifting dust is concentrated into an isolated annulus of terrestrial planetesimals. Gravitational scattering during terrestrial planet formation causes radial spreading, transporting planetesimals from inside 1-1.5 AU out to the belt. Several times the total current mass in S-types is implanted, with a preference for the inner main belt. C-types are implanted from the outside, as the giant planetsÕ gas accretion destabilizes nearby planetesimals and injects a fraction into the asteroid belt, preferentially in the outer main belt. These implantation mechanisms are simple byproducts of terrestrial- and giant planet formation. The asteroid belt may thus represent a repository for planetary leftovers that accreted across the Solar System but not in the belt itself.
\end{sciabstract}


\noindent {\bf \sc Introduction} \\
\indent According to disk models\cite{hayashi81,bitsch15}, the primordial asteroid belt contained at least an Earth mass in solids, a factor of more than 2000 higher than the belt's current total mass\cite{demeo13}. C-types -- which dominate the outer main belt\cite{gradie82,demeo13,demeo14} -- contain roughly three times more mass than S-types. The four largest asteroids, Ceres, Vesta, Pallas and Hygiea, contain more than half the belt's mass, and smaller asteroids contain a total of $\sim 2 \times 10^{-4} \mearth$\cite{demeo13}. Models have strived to explain the asteroid belt's mass deficit by by invoking dynamical depletion mechanisms such as sweeping secular resonances\cite{nagasawa00}, asteroidal planetary embryos\cite{petit01}, and Jupiter's orbital migration\cite{walsh11}. 

The present-day belt retains a memory of macroscopic bodies but not of primordial dust. Disk models may therefore not reflect the distribution of planetesimals. Indeed, the formation of planetesimals from dust is a complex process that may not occur everywhere in the disk\cite{chambers10,johansen14,armitage16}. Models show that drifting dust and pebbles are radially concentrated at pressure bumps in the disk\cite{haghighipour03}. Observations of the TW Hydra disk have found ringed substructure, a signature of particle drift\cite{andrews16}. This process can concentrate small particles and produce a narrow annulus of planetesimals in the terrestrial planet region that may not extend into the asteroid belt\cite{drazkowska16,surville16}.  Accretion from an annulus can match the terrestrial planets\cite{hansen09,walsh16} but it has been thought that additional mechanisms were needed to explain the asteroid belt.

Here we disprove the notion that the primordial belt must have been high-mass. We show that implantation of planetesimals into an empty asteroid belt can explain the total mass, orbital distribution and radial dichotomy of the present-day asteroids.

\noindent {\bf \sc Results} \\
\indent We performed a suite of simulations of terrestrial planet formation from an annulus in a dissipating gaseous disk. We included Jupiter and Saturn on low-eccentricity orbits, locked in 3:2 resonance with Jupiter at 5.4 AU. This configuration is consistent with their migration in the disk\cite{morby07a,pierens14} and the later evolution of the Solar System\cite{nesvorny12}. We built upon the {\tt Mercury}\cite{chambers99} symplectic integrator by implementing a prescription for aerodynamic gas drag\cite{adachi76,brasser07} and tidal damping\cite{cresswell07} within an underlying gas disk profile taken from hydrodynamical simulations\cite{morby07a} (Fig.~S1). The disk dissipated exponentially on a $2 \times 10^{5}$~yr timescale and was removed entirely after $2 \times 10^{6}$~yr. Simulations included $2-2.5 \mearth$ spread between 0.7 and either 1 or 1.5 AU, divided between 50-100 planetary embryos and a swarm of 2000-5000 100~km-sized planetesimals, with 75-90\% of the mass in embryos (see Supplementary Materials). 

\begin{figure}
\begin{center}
  \includegraphics[width=0.65\textwidth]{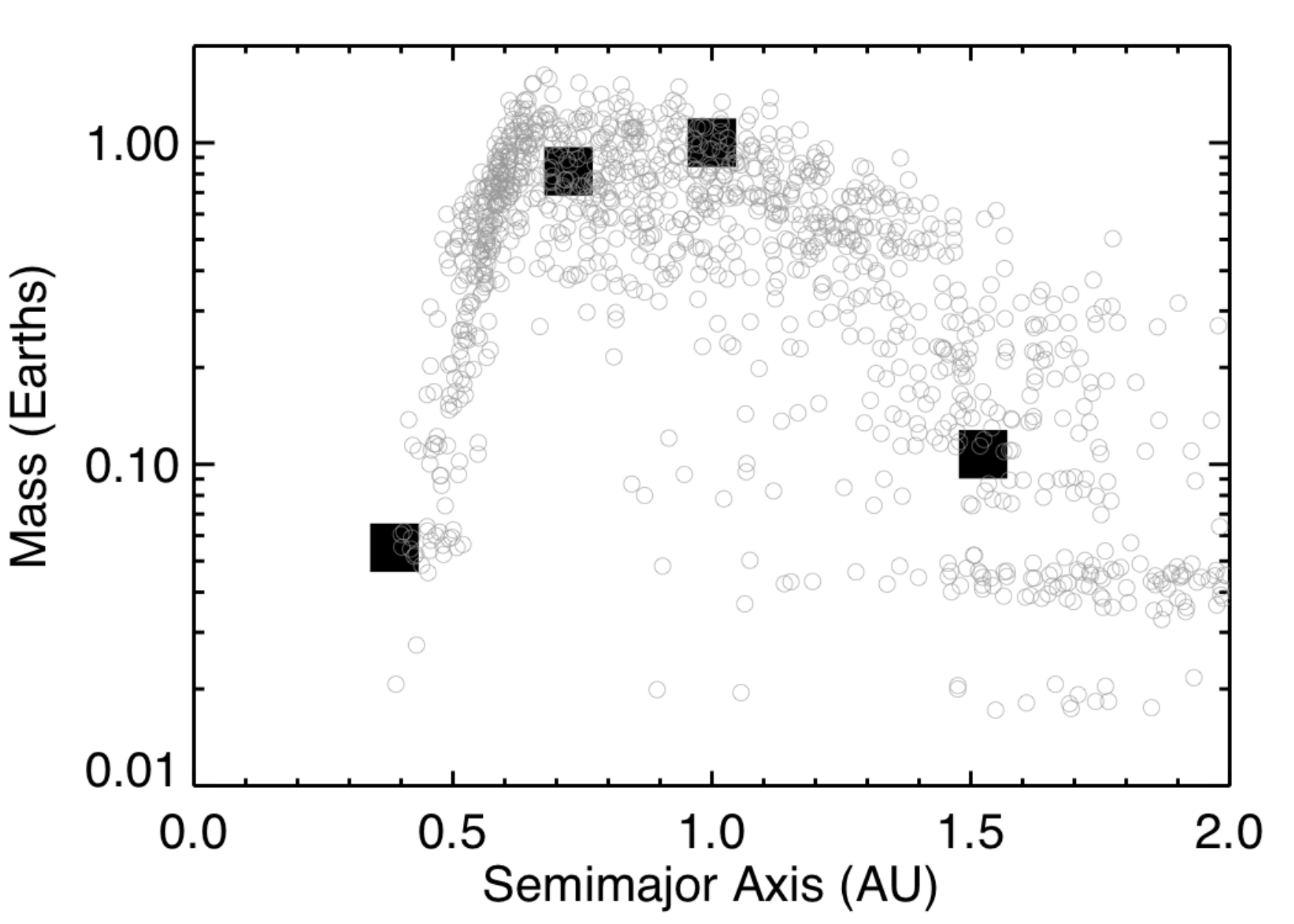}
 \includegraphics[width=0.65\textwidth]{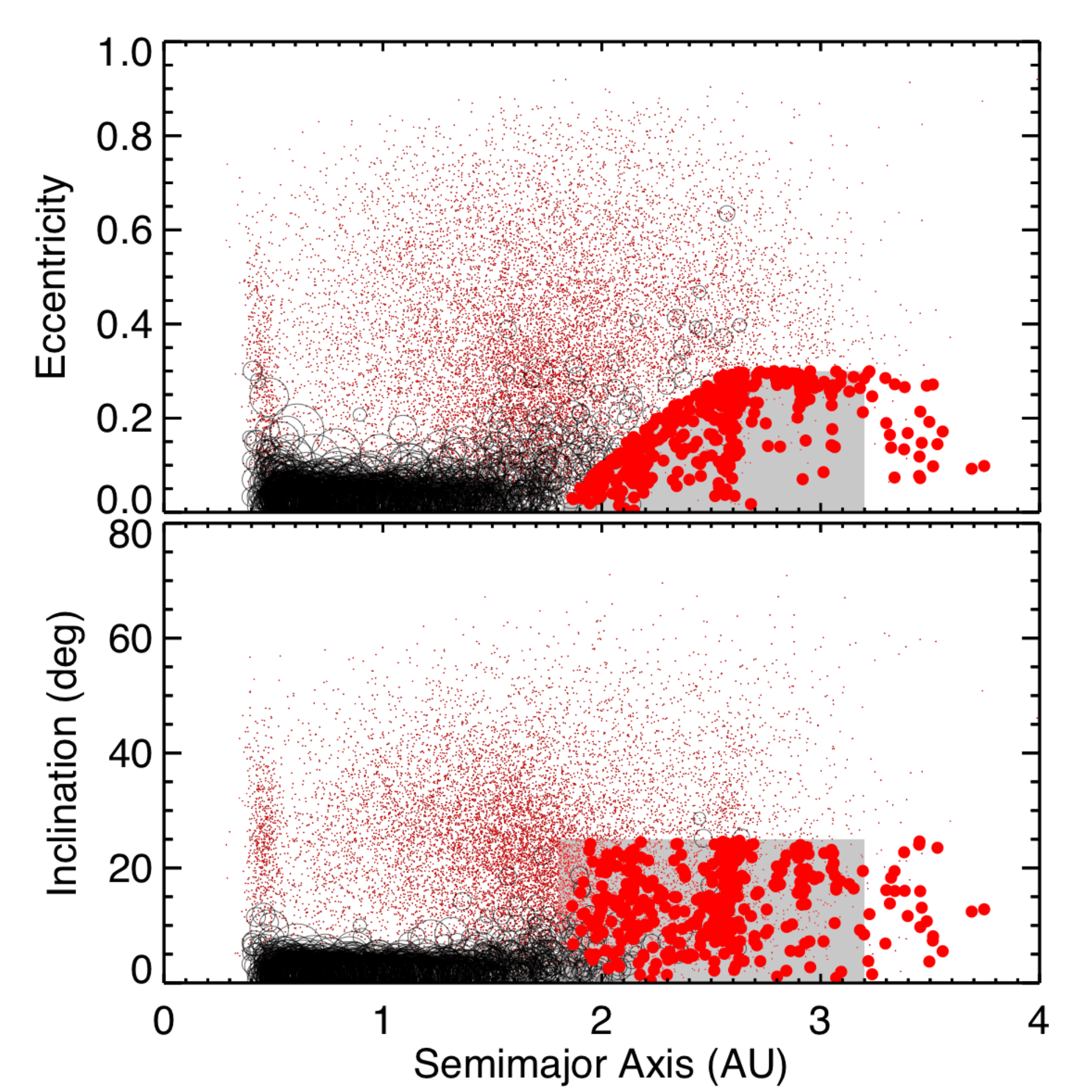}
    \caption[]{Outcome of our simulations. {\bf Top:} Semimajor axis-mass distribution of the terrestrial planets that formed in our simulations. The black squares represent the actual terrestrial planets. The clustering of planets at $\sim 0.05 \mearth$ simply represents the starting embryo mass. While some planets on Mars-like orbits were far larger than the actual one, we only included simulations with good Mars analogs in determining the abundance of implanted S-types.
    {\bf Bottom:} Semimajor axis-eccentricity and inclination distribution of S-type asteroids implanted from the terrestrial planet region. All planetesimals from the end of the simulations are shown, and the implanted ones are solid.  The shaded region represents the main asteroid belt, defined here as having perihelion distance $q > 1.8$~AU, eccentricity $e < 0.3$ and inclination $i < 25^\circ$.} 
     \label{fig:terrestrials}
     \end{center}
\end{figure}

During accretion, gravitational stirring caused the annulus of bodies to spread out\cite{hansen09,walsh16}. Mars may represent an embryo that was kicked out of the annulus and starved. Our simulations produced planets that broadly match the real terrestrial planets, with small Mars and Mercury analogs and large Earth and Venus analogs (Fig.~\ref{fig:terrestrials}). The orbital eccentricities and inclinations of our simulated terrestrial planetary systems were comparable to or slightly smaller than their current values. Given that their orbits would have been excited during the later giant planet instability\cite{kaib16,brasser13}, our simulated systems are consistent with the present-day terrestrial planets.

Planetesimals are often scattered by planetary embryos onto orbits that cross the asteroid belt. To be captured in the belt requires a planetesimal's perihelion distance to be lifted to avoid subsequent encounters with the growing terrestrial planets. This can happen either by scattering from a rogue embryo also on an asteroid belt-crossing orbit, a decrease in eccentricity driven by resonant interaction with the giant planets (typically at the 3:1 resonance with Jupiter at 2.6 AU), or a gas drag-driven decrease in eccentricity (although gas drag is not the main driver of capture; see Fig.~S2).

\begin{figure}
  \includegraphics[width=0.95\textwidth]{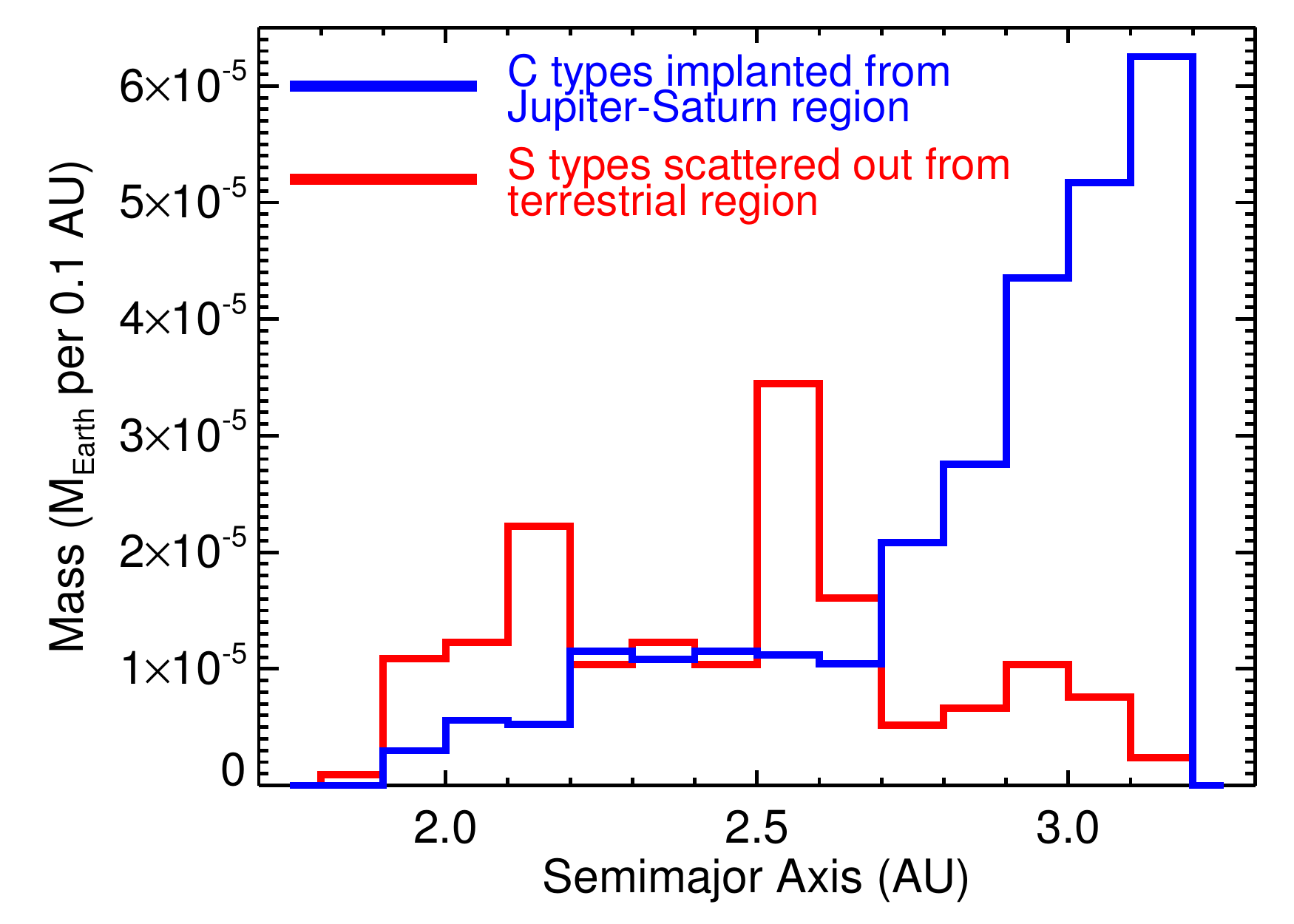}
    \caption[]{Radial distribution of implanted asteroids. S-types are 100~km planetesimals implanted onto stable orbits within the main belt from 159 simulations of terrestrial planet formation. The total mass in S-types was determined using only those simulations that matched Mars' mass and did not strand embryos in the asteroid belt\cite{raymond09c} but the distribution of all implanted asteroids was used, as we found no systematic difference. The C-types are 100~km planetesimals implanted during Jupiter and Saturn's rapid gas accretion\cite{raymond17}, calibrated to contain 1.7 times as much total mass as the S-types\cite{demeo13}.} 
     \label{fig:SCtypes}
\end{figure}

In our simulations planetesimals were implanted across the belt, with a preference for the inner main belt (Figs.~\ref{fig:terrestrials} and ~\ref{fig:SCtypes}). The peak in the distribution at 2.6 AU corresponds to Jupiter's 3:1 resonance, where objects are unlikely to be long-term stable\cite{gladman97}. However, if Jupiter's orbit were moving at the time (perhaps due to an early giant planet instability\cite{kaib16}) the resonant interaction could enhance the capture rate by implanting asteroids and then stranding as the resonance moves.

The total present-day mass in S-types is $3.8 \times 10^{-5} \mearth$\cite{demeo13}. The belt has lost a factor of a few in mass over the course of the Solar System's history, from the putative giant planet instability\cite{morby10} and long-term dynamical loss\cite{minton10}. Given the low probability of implantation and the long run-time of our simulations (1-4 months of CPU-time per simulation), we combine our simulations into a single distribution of outcomes. Our simulations implanted an average of 1.1 planetesimals onto stable orbits within the main belt. If we restrict ourselves to simulations that match the detailed characteristics of the Solar System, with a Mars analog within a factor of two of its actual mass and no embryos surviving in the asteroid belt\cite{raymond09c}, the rate drops to 0.68 trapped asteroids per simulation (note that all implanted asteroids were included in creating the distributions in Figs.~\ref{fig:terrestrials} and~\ref{fig:SCtypes} but only ``good'' simulations -- defined as those with Mars analogs smaller than twice its actual mass and no surviving embryos in the asteroid belt -- were used for calculating implantation efficiencies and masses). 

Our simulations implanted a mean of $1.6 \times 10^{-4} \mearth$ of terrestrial planetesimals into the belt (and up to ten times that amount in some simulations). This is more than four times the total mass in S-types. The mass in implanted planetesimals drops by $\sim 30$~\% if we make the extreme assumption that all implanted objects within 0.05 AU of Jupiter's 3:1 resonance at 2.6 AU will ultimately become unstable, but the implanted mass remains three times the total S-type mass. 

In our simulations planetesimals were implanted to the main belt from across the terrestrial region. The broad source region may explain the diversity among different types of asteroids in the inner parts of the belt\cite{gradie82,demeo13,demeo14}. However, planetesimals were implanted with a efficiency that depended on their starting location. The efficiency of implantation increased modestly in the Venus-Earth region, as planetesimals initially located from 0.9-1 AU had a $\sim$30\% higher efficiency of implantation than planetesimals initially located from 0.7-0.8 AU (Fig.~S3). Although only a subset of simulations started with planetesimals out to 1.5 AU (see Supplementary Materials), the efficiency of implantation increased dramatically beyond Earth's orbit. Planetesimals from the Mars region were implanted into the main belt with an efficiency more than ten times higher than for planetesimals starting near Venus' current orbit. Given Mars' small mass, the population of planetesimals initially located near Mars' current orbit was at most an order of magnitude smaller in total mass than the population of planetesimals near Earth and Venus' orbits.  However, if there was a primordial population of Mars-region planetesimals, its contribution to the present-day main belt may have been significant, and potentially comparable to that from the Earth-Venus region.

A separate mechanism can explain the origin of the C-types as planetesimals implanted from orbits exterior to the asteroid belt during Jupiter and Saturn's growth\cite{raymond17}. The C-types in Fig.~\ref{fig:SCtypes} were drawn from a simulation by Ref\cite{raymond17} in which Jupiter and Saturn's gas accretion destabilized nearby planetesimals' orbits. Planetesimals originating between roughly 4 and 9 AU were scattered by Jupiter and a fraction were implanted into the asteroid belt by the action of aerodynamic gas drag (see Fig.~S4). The disk's surface density profile and depletion rate match the simulations of S-type implantation. The efficiency of implantation of C-types into the main belt is significant but the amount of mass implanted depends on a number of unknown quantities such as the abundance of planetesimals in the primordial giant planet region, the detailed disk structure and the giant planets' growth and migration histories\cite{raymond17}. In Fig.~2 the implanted mass in C-types was calibrated to be 1.7 times the total mass in implanted S-types, the actual value when Ceres is removed\cite{demeo13}.

Implanted asteroids qualitatively match the observed S- vs C-type dichotomy\cite{gradie82,demeo14}. S-types dominate the main belt interior to $\sim 2.7$~AU and the C-types farther out (Fig.~\ref{fig:SCtypes}). Considering compositional variations within their broad source region, implantation may also explain the diversity in asteroid types and meteorite classes\cite{bus02,demeo14}. The inclination distribution of implanted S-types provides a reasonable match to the current ones. While the eccentricities of implanted S-types are modestly higher than present-day values, eccentricity re-shuffling during the giant planet instability\cite{morby10} should smooth out the distribution to match the current one~\cite{deienno16}.

Our simulations do not include collisional evolution, which grinds planetesimals down in time\cite{bottke05}. We account for this by calibrating the effective planetesimal mass to the ``late veneer'' inferred from highly-siderophile elements in Earth's mantle\cite{day07}.  Earth is thought to have accreted the final 0.5-1\% of its mass after the last giant impact\cite{morbywood15}.  If we restrict our simulations to roughly match the timing of the last giant impact, Earth analogs accreted a median of 0.4-0.8\% of an Earth mass, depending on the exact assumptions. This indicates that the simulated mass deposited in the asteroid belt is a fair representation of reality.  

\noindent {\bf \sc Discussion} \\
\indent The asteroid belt's size-frequency distribution (SFD) is a subject of vigorous study, as it constrains models of accretion and collisional evolution~\cite{bottke05,morby09,weidenschilling11}. There appear to be modest differences between the SFDs of S-types and C-types, and those differences may vary with orbital radius within the main belt~\cite{gladman09,masiero11,demeo13}. We have invoked two drastically different implantation mechanisms. S-types are dynamically injected and so should roughly preserve their source size distribution. However, they are likely to have undergone size-dependent collisional grinding before implantation~\cite{bottke06}. Gas drag-assisted capture of C-types is size-dependent\cite{raymond17}. The efficiency of implantation depends primarily on the strength of gas drag felt by planetesimals scattered by the growing giant planets, and this is a function of the disk properties (which vary in time) and the planetesimal size\cite{raymond17}. Simulations of the streaming instability find that planetesimals are likely to be born with a universal SFD regardless of where they form\cite{johansen15,simon16}. Given the size-dependence of the processes involved, the existence of compositional and spatial variations in the SFD among implanted asteroids is to be expected.

Additional mechanisms may contribute to producing `refugee' asteroids. Stochastic forcing from MHD turbulence in the disk can generate radial excursions of planetesimals\cite{nelson05}. If Jupiter's core formed interior to Mercury's present-day orbit and migrated outward it would have transported some planetesimals from the inner Solar System to the main belt\cite{raymond16}. Ref \cite{bottke06} proposed that the asteroid Vesta, as well as the parent bodies of iron meteorites, were scattered outward from the terrestrial planet region.  However, the simulations from Ref\cite{bottke06} included far too much mass in the Mars region and asteroid belt and were not consistent with the large Earth/Mars mass ratio\cite{raymond09c}. Nonetheless, it is interesting to note that the mechanism of S-type implantation from the terrestrial planet region operates even if the primordial asteroid belt was not empty, and actually with a slightly higher implantation efficiency than the one we find in our simulations with an empty primordial belt\cite{bottke06}. This implies that, regardless of the formation scenario, remnants from the Earth-Venus region must exist in the belt.

Asteroidal implantation solves a problem for the model of Solar System formation in which the terrestrial planets formed from a narrow annulus\cite{hansen09,walsh16,drazkowska16,morbyraymond16}.  As we have shown, the primordial empty asteroid belt is populated with S-types from the inside as a simple consequence of gravitational spreading during terrestrial planet formation. External implantation is required to explain the C-types, and this happens as a simple consequence of gas accretion onto the growing giant planets\cite{raymond17}. Late in the disk phase, planetesimals are scattered by the same mechanism past the asteroid belt to deliver water to the terrestrial planets\cite{raymond17}. The mechanisms of both S-type and C-type implantation are largely independent of the giant planets' formation and migration, and are thus consistent with a wide range of giant planet formation models\cite{bitsch15b,levison15}.

Our analysis does not prove that no asteroids formed in the belt. Rather, the total mass and large-scale distribution of S-types is reproduced as a byproduct of terrestrial planet formation from an annulus. Rather than invoking a large mass in asteroids that requires later depletion, it is worth considering that the primordial asteroid belt may have simply been empty.  If this is true, the belt may represent a cosmic refugee camp, a repository for planetesimals implanted from across the Solar System, none of which calls the asteroid belt home.

\vskip .4in
\noindent {\bf \sc Materials and Methods} \\
\indent Our code is based on the {\tt Mercury} integration package\cite{chambers99}.  We added synthetic forces to mimic gas-disk interactions.  As our simulations start after Jupiter and Saturn are assumed to have formed, we included an underlying disk profile (Fig.~\ref{fig:gas}) drawn from hydrodynamical simulations of the giant planets embedded in the gaseous protoplanetary disk\cite{morby07a}.

\setcounter{figure}{0} 
\renewcommand{\thefigure}{S\arabic{figure}}
\begin{figure}[h]
\begin{center}
  \includegraphics[width=0.65\textwidth]{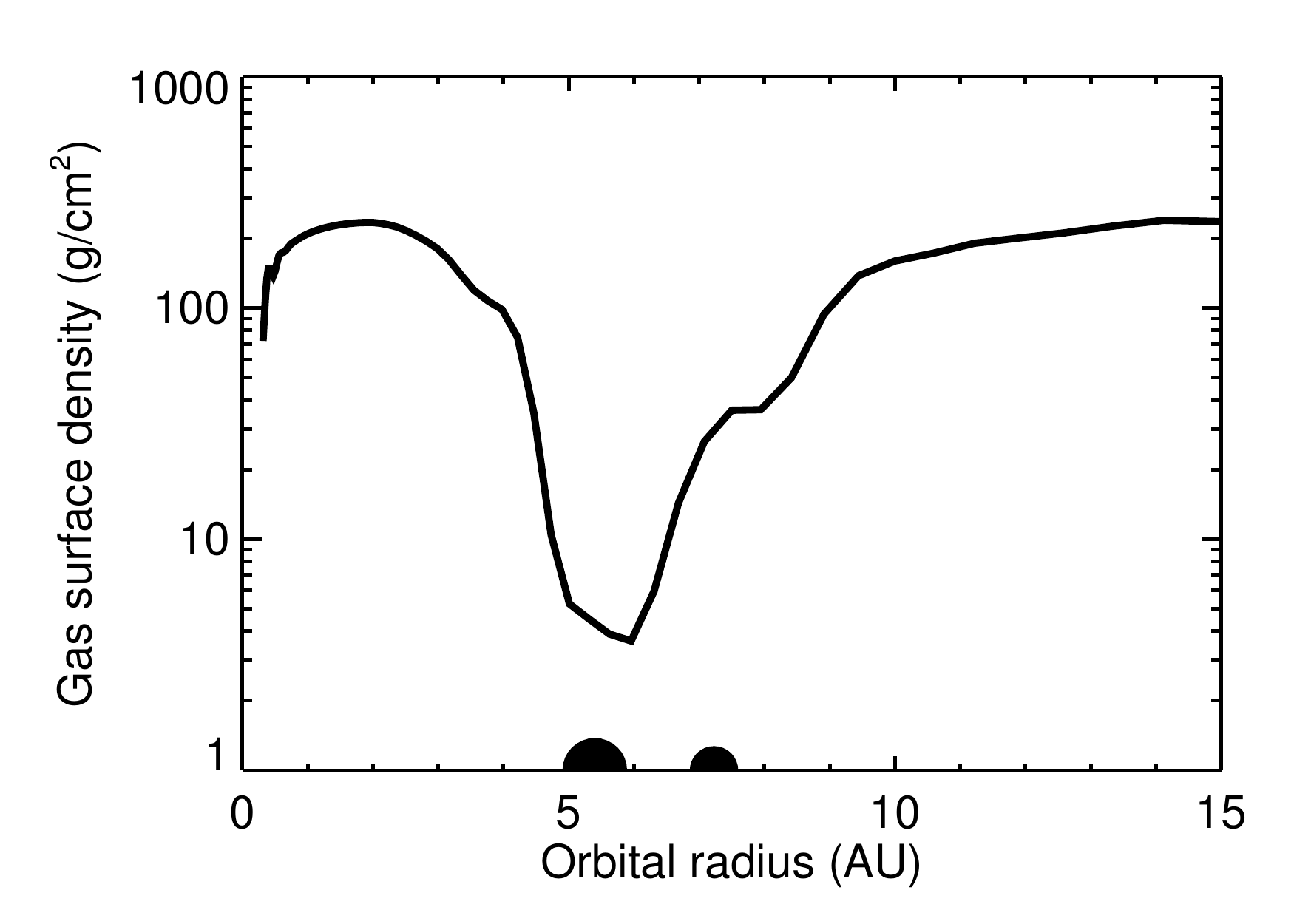}
    \caption[]{Surface density profile of the underlying gas disk profile used in our simulations, drawn from hydrodynamical simulations\cite{morby07a}. The positions of Jupiter and Saturn are included at the bottom.  This is the starting disk profile.  In  simulations the surface density was decreased uniformly in radius on a $2 \times 10^5$~yr exponential timescale and removed entirely after 2 Myr. } 
     \label{fig:gas}
\end{center}
\end{figure}

For planetesimal particles we calculated an additional acceleration from aerodynamic gas drag, defined as: 
\begin{equation}
{\rm {\bf a}_{drag} =- \frac{3C_d \rho_g v_{rel} \bf{v}_{rel}}{8 \rho_p R_p}}
\end{equation}
where ${\rm C_d}$ is the drag coefficient,  ${\rm \rho_p}$ and ${\rm R_p}$ are the planetesimal's bulk density (fixed at $1.5 \, g\, cm^{-3}$) and size, respectively. We fixed the planetesimal size at 100~km, i.e., $R_p = 50$~km. The ${\rm v_{rel}}$ vector is the relative velocity of the object with respect to the surrounding gas and ${\rm  \rho_g}$ is the gas density at the planetesimal location. The gas drag coefficient $C_d$ is implemented following\cite{brasser07}.  

We also included the effect of type 1 damping on the embryos' eccentricities and inclinations\cite{papaloizou00,tanaka04}. The damping time scale is defined as\cite{tanaka04}
\begin{equation}
t_{wave} = \frac{1}{\Omega_p}\frac{M_\star}{m_p} \frac{M_\star}{\Sigma_p a^2_p} \left(\frac{H}{r}\right)^4,
\end{equation}
where $\Omega_p$ is the orbital angular velocity, $M_\star$ and $m_p$ are the stellar and planetary mass, respectively, $\Sigma_p$ is the local disk surface density, $a_p$ is the planet's semimajor axis, and $H/r$ the local disk aspect ratio.  

The eccentricity damping timescale $t_e$ is \cite{cresswell07}:
\begin{equation}
t_e = \frac{t_{wave}}{0.78}\left[1 - 0.14\left(\frac{e}{H/r}\right)^2 + 0.06 \left(\frac{e}{H/r}\right)^3 + 0.18 \left(\frac{e}{H/r}\right) \left(\frac{i}{H/r}\right)^2\right]
\end{equation}
and the inclination damping timescale $t_i$ is
\begin{equation}
t_i = \frac{t_{wave}}{0.544}\left[1 - 0.30\left(\frac{i}{H/r}\right)^2 + 0.24 \left(\frac{i}{H/r}\right)^3 + 0.14 \left(\frac{e}{H/r}\right)^2 \left(\frac{i}{H/r}\right)\right].
\end{equation}
Given the small masses of embryos in the simulations ($\sim 0.04 \mearth$), radial migration was too slow to have an effect.

Our simulations started from a narrow annulus of terrestrial building blocks, as this has been shown to quantitatively reproduce the radial mass distribution and orbital excitation of the terrestrial planets \cite{hansen09,walsh16}. In all cases the inner edge was fixed at 0.7 AU.  We ran nine sets of simulations with slightly different initial distributions of terrestrial material (see Supplementary Materials for details).  All of them contained $2-2.5 \mearth$ divided between 50-100 planetary embryos and either 2000 or 5000 planetesimals, initially laid down with an inner edge at 0.7 AU and an outer edge at 1 AU or 1.5 AU. 

Embryos and planetesimals were given small, randomized non-zero eccentricities of up to 0.02 and inclinations of up to $1^\circ$.  Simulations were run for 200 million years with a 10-day timestep, sufficient to accurately resolve orbits in to a few tenths of an AU\cite{rauch99,levison00,raymond11}.  We inflated the star's radius to 0.2 AU to avoid numerical error.  Objects were considered ejected if they reached a heliocentric distance of 100~AU. 

Of the 280 simulations, 273 ran to completion and were included in the main paper. We consider 98 of the simulations to have ÒgoodÓ outcomes in terms of MarsÕ mass (with a Mars analog less than twice MarsÕ actual mass) and with no embryos stranded beyond Mars\cite{raymond09c}.


\begin{scilastnote}
\item We thank the Agence Nationale pour la Recherche for support via grant ANR-13-BS05-0003-002 (grant MOJO).  A. I. thanks financial support from FAPESP (Process numbers: 16/12686-2 and 16/19556-7). S.~N.~R. also thanks NASA Astrobiology InstituteÕs Virtual Planetary Laboratory Lead Team, funded via the NASA Astrobiology Institute under solicitation NNH12ZDA002C and Cooperative Agreement Number NNA13AA93A. The authors declare that they have no competing interests. All of our simulations are fully available upon request (email S.~N.~R.). This project proceeded as follows: it was instigated by S.~N.~R.; the simulation code was developed by A.~I.; S.~N.~R. ran and analyzed the simulations, and wrote the paper with considerable input from A.~I.
\end{scilastnote}

\pagebreak
\section*{Supplementary Materials}

In this Supplement we 1) detail and justify our initial conditions, 2) present some details of the timing, pathway and efficiency of asteroidal implantation, and 3) present the mechanism of C-type asteroid implantation during the giant planets' growth \cite{raymond17}.

\vskip .2in
\noindent {\bf 1. Initial conditions}

The main simulations started from a narrow annulus of terrestrial building blocks, as this has been shown to quantitatively reproduce the radial mass distribution and orbital excitation of the terrestrial planets\cite{hansen09,walsh16}.  In all cases the inner edge was fixed at 0.7 AU.

We ran sets of simulations with slightly different assumptions:
\begin{itemize}
\item 50 simulations with an outer edge at 1 AU that included 50 embryos totaling $2 \mearth$ and 2000 planetesimals totaling $0.5 \mearth$, laid out to follow an $r^{-1}$ surface density profile.
\item 15 simulations with 50 embryos totaling $2 \mearth$ and 2000 planetesimals totaling $0.5 \mearth$, laid out to follow an $r^{-1}$ surface density profile.  Twenty percent (400) of the planetesimals and embryos (10) were spread out between 1 and 1.5 AU.
\item 30 simulations with 50 embryos totaling $1.6 \mearth$ and 2000 planetesimals totaling $0.6 \mearth$, laid out to follow an $r^{-1}$ surface density profile.  Twenty percent (400) of the planetesimals and embryos (10) were spread out between 1 and 1.5 AU.
\item 20 simulations with 50 embryos totaling $2 \mearth$ and 2000 planetesimals totaling $0.5 \mearth$, laid out to follow an $r^{-5.5}$ surface density profile out to 1.5 AU\cite{izidoro15c}.
\item 30 simulations with 50 embryos totaling $1.6 \mearth$ and 2000 planetesimals totaling $0.6 \mearth$, laid out to follow an $r^{-1}$ surface density profile out to 1 AU.
\item 30 simulations with 50 embryos totaling $2 \mearth$ and 2000 planetesimals totaling $0.2 \mearth$, laid out to follow an $r^{-1}$ surface density profile out to 1 AU.
\item 20 higher-resolution simulations run with the same gas disk as the main simulations.  They included 100 embryos totaling $1.5 \mearth$ and 5000 planetesimals totaling $0.5 \mearth$ following an $r^{-1}$ surface density profile from 0.7-1 AU.  These simulations each took 3-4 months to run on a dedicated core, as compared with roughly one month each for the lower-resolution simulations. 
\item 50 simulations with an outer edge at 1 AU that included 50 embryos totaling $2 \mearth$ and 2000 planetesimals totaling $0.5 \mearth$, laid out to follow an $r^{-1}$ surface density profile, in which planetesimals were assumed to be 10~km in radius in terms of the gas drag calculation.
\item 35 simulations without the effects of the gaseous disk.  These included 1) 20 high-resolution simulations with 100 embryos totaling $1.5 \mearth$ and 5000 planetesimals totaling $0.5 \mearth$ following an $r^{-1}$ surface density profile from 0.7-1 AU; and 2) 15 simulations with an outer edge at 1 AU that included 50 embryos totaling $2 \mearth$ and 2000 planetesimals totaling $0.5 \mearth$, laid out to follow an $r^{-1}$ surface density profile.
\end{itemize}

Of the 280 simulations, 273 ran to completion and were included in the main paper. We consider 98 of the simulations to have ``good'' outcomes in terms of Mars' mass (with a Mars analog less than twice Mars' actual mass) and with no embryos stranded beyond Mars\cite{raymond09c}.

\vskip .2in
\noindent {\bf 2. Additional details about S-type Implantation}

\begin{figure}
\begin{center}
  \includegraphics[width=0.49\textwidth]{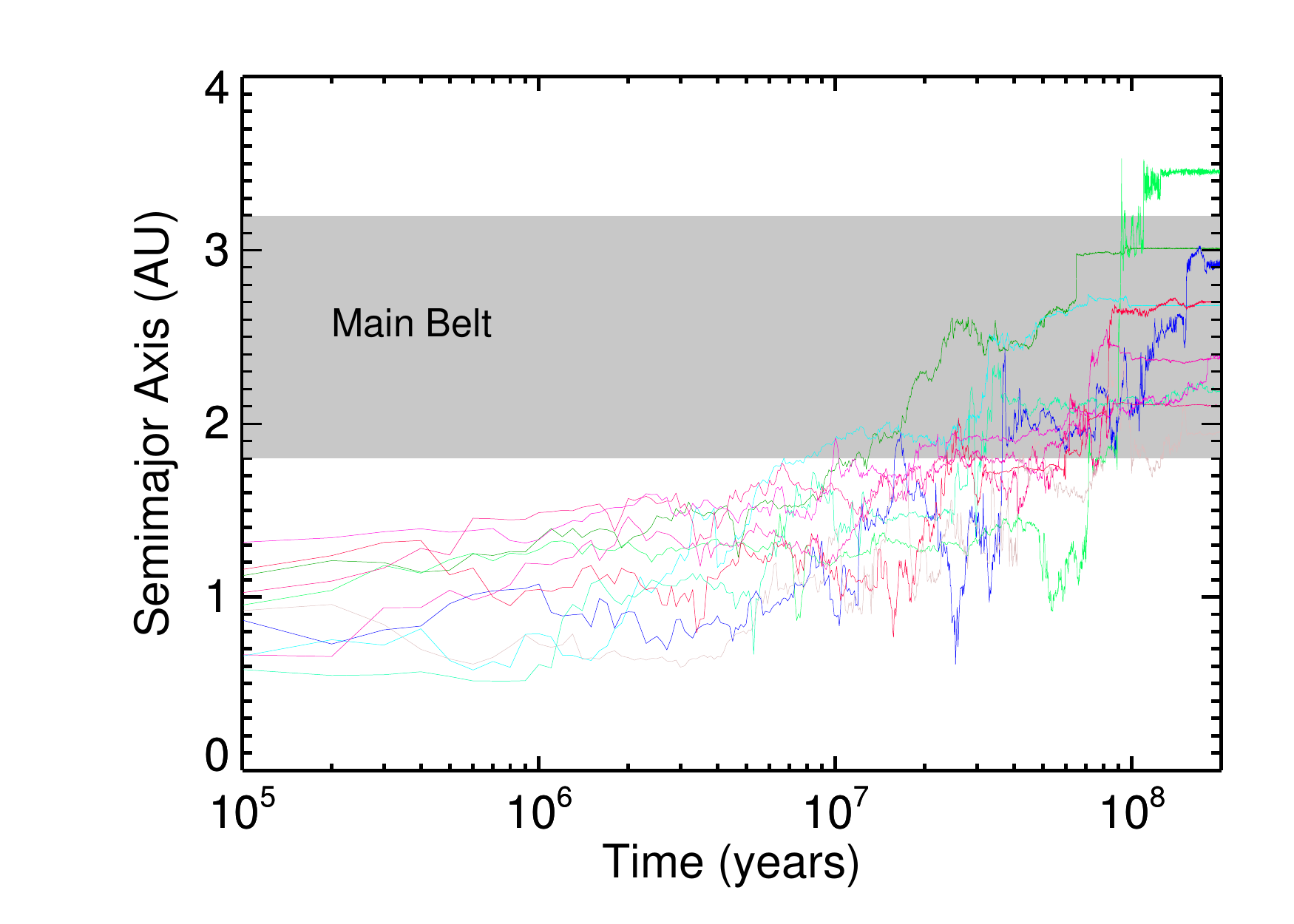}
    \includegraphics[width=0.49\textwidth]{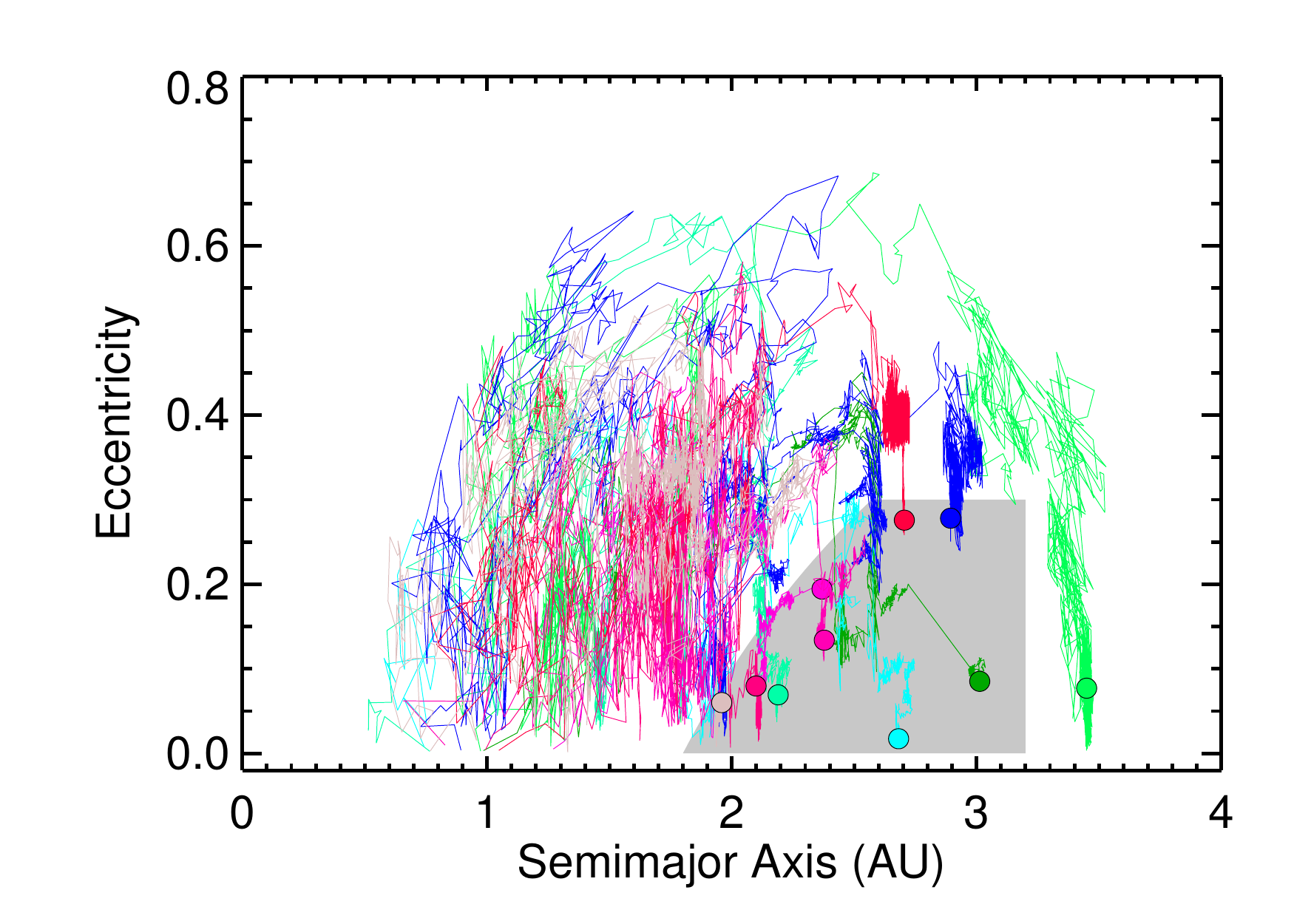}
    \caption[]{How ten asteroids from different simulations were implanted into the belt. {\bf Left:} Time evolution of the semimajor axis for the asteroids.  Each asteroid's path is traced in a unique color.
    {\bf Right:} Evolution of the same ten asteroids (with the same colors) in semimajor axis-eccentricity space.  The shaded region corresponds to the main belt. The solid points are the final orbits at the end of the simulations.
    } 
     \label{fig:implant}
     \end{center}
\end{figure}

Figure~\ref{fig:implant} shows the timing and orbital pathway of 10 representative asteroids that were implanted from the terrestrial planet region into the asteroid belt.  The key points are the following.  First, asteroids are only scattered to the asteroid region after the gas disk has dissipated.  The median implanted asteroid was scattered onto a belt-crossing orbit after 9 Myr, and was implanted into the belt after roughly 60 Myr. This explains why the presence of the gas disk in our simulations has a small effect on the implantation rate. It is also reassuring in terms of our initial conditions.  Our initial placement of embryos was very compact and borderline unstable.  However, the fact that implantation happens later, after the initial conditions were `forgotten'\cite{kokubo06}, means that the exact initial distribution of embryos is probably not important. Second, asteroids follow a complicated path in semimajor axis-eccentricity space.  They spend a significant amount of time at high-eccentricity and typically enter the main belt as the result of an eccentricity-decreasing event, typically by being scattered by a rogue embryo that is on a path toward an encounter with Jupiter and subsequent ejection from the system.

Figure~S3 shows the efficiency of implantation for planetesimals as a function of their starting orbital radius.  While implanted asteroids sample the entire terrestrial disk, their is a significant dependence on the initial position.  Within the Venus-Earth annulus from 0.7-1 AU, the implantation efficiency increases modestly with radius and there is a $\sim 30\%$ difference between the efficiency of implantation for the 0.7-0.8 AU and 0.9-1 AU bins.  The efficiency rises dramatically for planetesimals initially located beyond 1 AU, and planetesimals initially located between 1.25 and 1.5 AU are implanted at an efficiency that is more than ten times higher compared with planetesimals initially from 0.7 to 0.8 AU.  

Compared with objects initially within 1 AU, planetesimals initially located beyond 1 AU were implanted at a modestly higher rate into the outer main belt past 2.6 AU: 107 of 321 implanted planetesimals (33\%) from $r_{init} < 1$~AU were implanted into the outer main belt vs 20 of 47 (43)\% for $r_{init} > 1$~AU.  In addition, planetesimals from beyond 1 AU were implanted modestly earlier than those from within 1 AU, with a median implantation time 10 million years earlier (median of 56 vs 66 Myr).  However, given the relatively small number of implanted planetesimals -- in particular from beyond 1 AU -- these last two correlations are not statistically significant and will require further simulations to confirm. 

Implantation requires a series of scattering events to put the planetesimal onto a belt-crossing orbit, and a mechanism to decrease the planetesimal's eccentricity and trap it on a stable orbit.  Of course, if a planetesimal's eccentricity can be decreased -- either by a rogue embryo or resonant interaction with Jupiter or Saturn -- then it can also be increased and later de-stabilized.  Stable implantation thus occurs most readily when the process cannot be replayed in reverse, for example, if a rogue embryo scatters a planetesimal into the belt and is then ejected, or Jupiter's 3:1 resonance acts to drop a planetesimal's eccentricity and then Jupiter's orbit shifts, stranding the planetesimal on a non-resonant orbit within the main belt.

\begin{figure}
\begin{center}
  \includegraphics[width=0.65\textwidth]{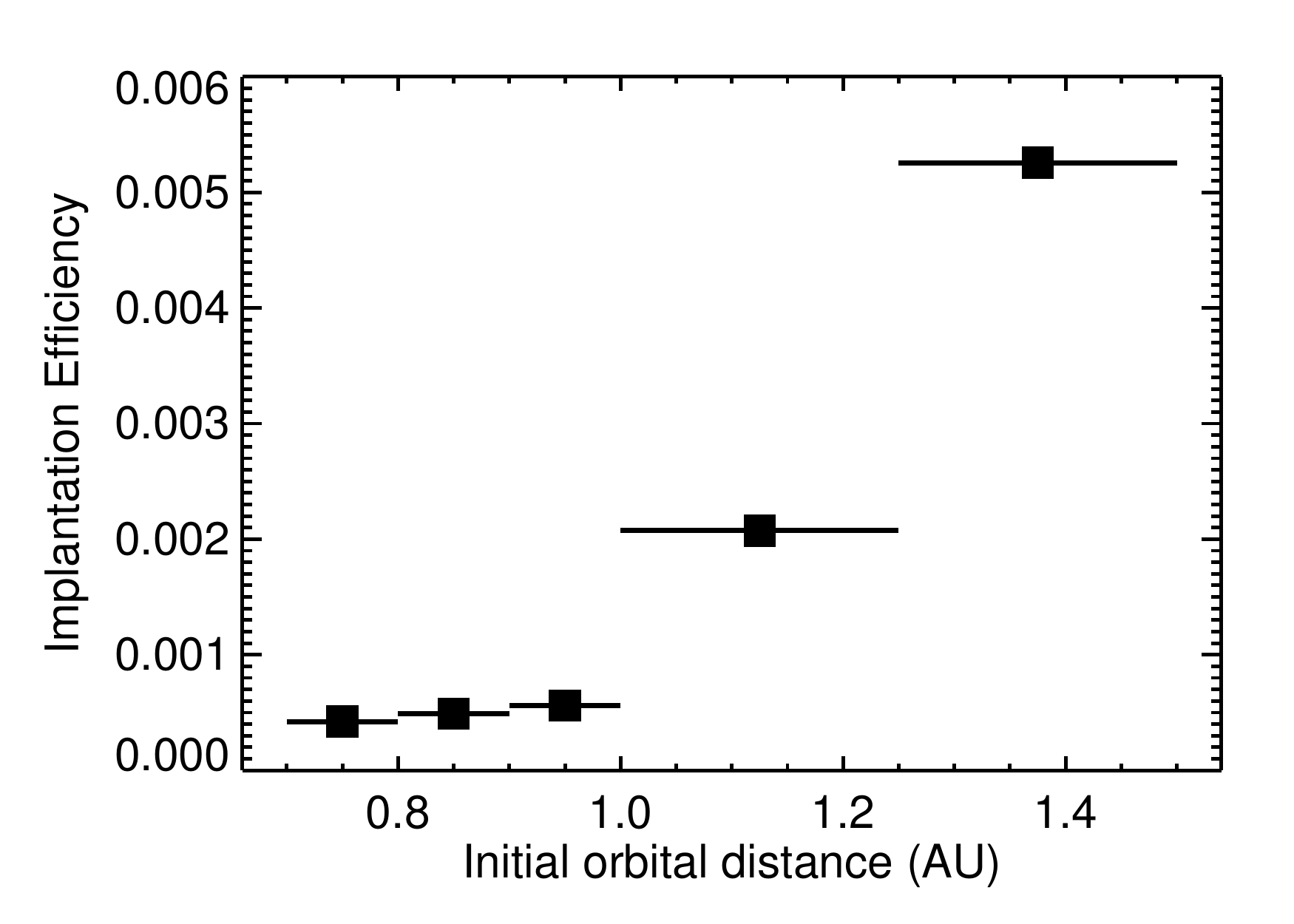}
    \caption[]{Efficiency with which planetesimals were implanted into the main asteroid belt in our simulations. Our initial parameter space was divided into five bins: 0.7-0.8 AU, 0.8-0.9 AU, 0.9-1 AU, 1-1.25 AU, and 1.25-1.5 AU.  Only a subset of simulation included planetesimals past 1 AU. 
    } 
     \label{fig:imp_eff}
     \end{center}
\end{figure}

Given the importance of close encounters, one would expect efficiency to be more effective when there is more mass in embryos in or near the belt.  Indeed, one motivation for our study is that Bottke et al's \cite{bottke06} simulations -- proposing that the parent bodies of iron meteorites were scattered out from the terrestrial planet region -- included far too much mass in the Mars region and asteroid belt to be consistent with Mars' actual mass.  While our simulations started with very little mass in the Mars region and an empty asteroid belt, we see a correlation between the mass of planets that accreted at Mars' distance and the implantation rate of planetesimals.  Simulations with a Mars larger than twice its actual mass or an embryo stranded in the belt (rogue or not, but lasting for 200 Myr) had an implantation rate twice as high as simulations with acceptable Mars analogs and no asteroidal embryos.  The distribution of planetesimals implanted in simulations with asteroidal embryos or too-massive Mars analogs also had a smaller peak at Jupiter's 3:1 resonance ($\sim$10\% of implanted population in/near the resonance vs. $\sim 30$\%).  Planetesimals were also implanted onto somewhat lower eccentricity orbits.  The statistical differences between the distributions depend on the cutoff for good vs. bad Mars analogs.

Nonetheless, the total mass in implanted planetesimals in the `good' simulations is sufficient to explain the S-types, and could reasonably be a factor of a few higher if there was a significant population of planetesimals in the Earth-Mars region. Additional processes could also have helped capture planetesimals such as moving resonances due to chaos in the giant planets' orbits\cite{izidoro16} or an early giant planet instability\cite{kaib16}.  An early instability might temporarily scatter an ice giant onto an asteroid belt-crossing orbit, which would smear out implanted planetesimals' eccentricities and inclinations\cite{brasil16}.  

\noindent {\bf 3. Implantation of C-type asteroids driven by the gas giants' growth}

In a separate paper\cite{raymond17} we presented a new mechanism for the implantation of planetesimals from the giant planet region into the asteroid belt. The implantation is driven by gas accretion onto the growing Jupiter and Saturn, which changes both the dynamical environment of nearby planetesimals (often destabilizing them) and the structure of the gas disk by carving an annular gap\cite{morby07a}.

\begin{figure}
\begin{center}
  \includegraphics[width=0.65\textwidth]{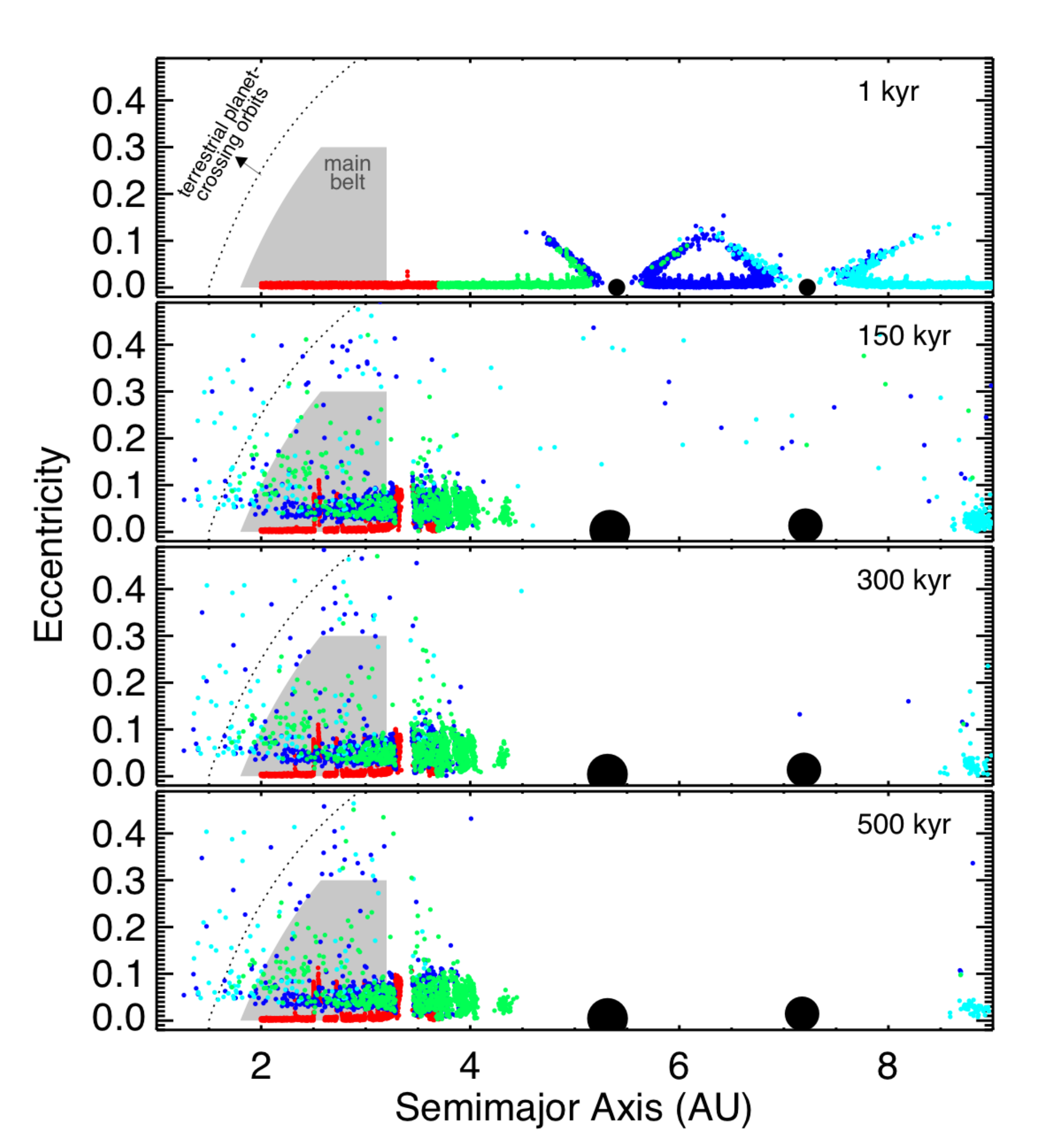}
    \caption[]{Snapshots in the evolution of a simulation in which the giant planets' growth implants planetesimals from the Jupiter-Saturn region as C-type asteroids.  Planetesimals are colored according to their starting orbital radius.  The main asteroid belt is shaded, and the first panel shows the region of parameter space (at high eccentricity and low semimajor axis) that crosses the terrestrial planets' orbits. Adapted from Ref \cite{raymond17}.
     } 
     \label{fig:ctypes}
     \end{center}
\end{figure}

Figure~\ref{fig:ctypes} illustrates this mechanism (in this case neglecting giant planet migration).  In this simulation, Jupiter grew from a $3 \mearth$ core to its current mass from 200 to 300 kyr, and Saturn grew from 300 to 400 kyr.  The structure of the underlying gaseous disk was interpolated between an $r^{-1}$ surface density profile and a disk with a single gap carved by Jupiter, and finally the more complex profile shown in Fig.~\ref{fig:gas} as Saturn grows\cite{morby07a}. Planetesimals were assumed to be 100~km in diameter for the gas drag calculation\cite{adachi76}.

Figure~\ref{fig:ctypes} shows how the giant planets' phase of rapid gas accretion scatters nearby planetesimals.  A significant fraction of planetesimals initially between 4 and 9 AU were implanted into the main belt, preferentially in the outer part (see Fig.~\ref{fig:SCtypes}).  Many planetesimals were scattered past the main belt toward the terrestrial region, representing a source of water for the growing terrestrial planets.

The mechanism illustrated in Fig.~\ref{fig:ctypes} is robust to a number of parameters including the disk's dissipation rate and the giant planets' growth timescale and migration history.  For a full description of the mechanism, see Ref\cite{raymond17}.


\end{document}